\documentclass[twocolumn,showpacs,pre,superscriptaddress]{revtex4}

\usepackage{graphicx}% Include figure files
\usepackage{dcolumn}% Align table columns on decimal point
\usepackage{bm}% bold math
\usepackage{color}
\usepackage{amsmath}
\usepackage{sidecap}

\begin{document}

\title{Unifying model of driven polymer translocation}

\author{T. Ikonen}
\affiliation{Department of Applied Physics and COMP Center of
Excellence, Aalto University School of Science,
P.O. Box 11000,
FI-00076 Aalto, Espoo, Finland}

\author{A. Bhattacharya}
\affiliation{Department of Physics, University of Central Florida, Orlando, Florida 32816-2385, USA}

\author{T. Ala-Nissila}
\affiliation{Department of Applied Physics and COMP Center of
Excellence, Aalto University School of Science,
P.O. Box 11000,
FI-00076 Aalto, Espoo, Finland}
\affiliation{Department of Physics,
Box 1843, Brown University, Providence, Rhode Island 02912-1843}

\author{W. Sung}
\affiliation{Department of Physics, Pohang University of Science and Technology, Pohang 790-784, South Korea}

\date{\today}

\begin{abstract}
We present a Brownian dynamics model of driven polymer translocation, in which non-equilibrium memory effects arising from tension propagation (TP) along the {\it cis} side subchain are incorporated as a time-dependent friction. To solve the effective friction, we develop a finite chain length TP formalism, based on the idea suggested by Sakaue [Sakaue, PRE 76, 021803 (2007)]. We validate the model by numerical comparisons with high-accuracy molecular dynamics simulations, showing excellent agreement in a wide range of parameters. Our results show that the dynamics of driven translocation is dominated by the non-equilibrium TP along the {\it cis} side subchain. Furthermore, by solving the model for chain lengths up to $10^{10}$ monomers, we show that the chain lengths probed by experiments and simulations are typically orders of magnitude below the asymptotic limit. This explains both the considerable scatter in the observed scaling of translocation time w.r.t. chain length, and some of the shortcomings of present theories. Our study shows that for a quantitative theory of polymer translocation, explicit consideration of finite chain length effects is required.
\end{abstract}

\maketitle

\section{Introduction}

The transport of a polymer across a nanopore is vital to many biological processes, such as DNA and RNA translocation through nuclear pores, protein transport across membrane channels and virus injection~\cite{albertsbook}. Due to various potential technological applications such as rapid DNA sequencing, gene therapy and controlled drug delivery~\cite{meller2003}, polymer translocation has received considerable experimental~\cite{meller2003, kasi1996, storm2005} and theoretical interest~\cite{chuang2001,kantor2004, sung1996, muthu1999, dubbeldam2007,vocks2008,sakaue2007,sakaue2008,sakaue2010,saito2011, rowghanian2011,luo2008,luo2009,milchev2011,huopaniemi2006,bhatta2009,bhatta2010,metzler2010,huopaniemi2007,lehtola2008,lehtola2009,gauthier2008a,gauthier2008b,dubbeldam2011}. Of particular technological interest is the case of driven translocation, where the process is facilitated by an external driving force. The key theoretical issue here is to find a unifying physical description that yields the correct dynamical behavior, e.g., the dependence of the translocation time $\tau$ on the chain length $N_0$. Experiments and numerical simulations have indicated that $\tau \propto N_0^\alpha$. However, numerous different values of $\alpha$ have been observed, suggesting explicit dependence on the various physical parameters (cf. Ref.~\cite{milchev2011} for a recent review). Several theories of driven polymer translocation have emerged~\cite{vocks2008,dubbeldam2007,sakaue2007,sakaue2008,sakaue2010,saito2011,rowghanian2011,kantor2004,sung1996,muthu1999,dubbeldam2011}, some claiming agreement with the experimental or numerical results within a certain subset of the physical parameter space. However, to date no single theory has been able to capture the wide range of observed values of $\alpha$, nor quantitatively explain the reason for their dependence on the system's parameters. Therefore, the need for a unifying theory of driven translocation remains.

In Refs.~\cite{sung1996,muthu1999}, polymer translocation was described as a one-dimensional barrier crossing problem of the translocation coordinate $s$ (the length of the subchain on the {\it trans} side). Here, the chain starts from the {\it cis} side with one end inside the pore ($s=0$) and is considered as translocated once $s=aN_0$, with $a$ the segment length. The free-energy due to chain entropy and the chemical potential difference $\Delta\mu$ is
$\mathcal{F}(s)=(1-\gamma')k_BT\ln\left[\frac{s}{a}\left(N_0-\frac{s}{a}\right)\right] +\frac{s}{a}\Delta\mu.$
Here $\gamma'$ is the surface exponent ($\gamma'=0.5,~\approx 0.69,~\approx 0.95$ for an ideal chain, self-avoiding chain in 2D and 3D, respectively), and $k_BT$ is the thermal energy. From $\mathcal{F}(s)$, the Brownian dynamics equation for $s$ in the overdamped limit follows as
$\Gamma\frac{ds}{dt}=(1-\gamma')k_BT\left[ \frac{1}{aN_0-s} -\frac{1}{s} \right] - \frac{\Delta\mu}{a} + \zeta(t).$
Here $\Gamma$ is the (constant) effective friction, and $\zeta(t)$ is Gaussian white noise satisfying $\langle \zeta(t) \rangle=0$ and $\langle \zeta(t)\zeta(t') \rangle = 2\Gamma k_BT\delta(t-t')$.  For moderate to large $\Delta\mu$, this model describes translocation at constant mean velocity $\langle ds/dt \rangle = -\Delta\mu / a\Gamma$. However, it is known that the translocation process initially slows down and finally speeds up towards the end~\cite{storm2005, gauthier2008a, gauthier2008b, huopaniemi2006, lehtola2008, lehtola2009}. Qualitatively, this observation has been explained by a simple force-balance argument, where the friction $\Gamma$ depends on the number of moving monomers on the {\it cis} side subchain~\cite{storm2005, lehtola2008, lehtola2009,gauthier2008a, gauthier2008b}.  In Refs.~\cite{storm2005,gauthier2008a, gauthier2008b}, it is assumed that the whole subchain on the {\it cis} side is set into motion immediately after the force at the pore is applied. However, this assumption is only valid in the limit of extremely weak driving force, where the {\it cis} side subchain is always at equilibrium. %~\cite{sakaue2007, sakaue2010}.
In most cases, the driving force is substantially larger, implying that even the subchains are out of equilibrium~\cite{lehtola2008, lehtola2009, bhatta2009, bhatta2010}. It has been proposed that in this regime, the out-of-equilibrium dynamics can be described by tension propagation (TP) along the chain backbone, which leads to nontrivial time-dependence of the drag force and gives the non-monotonic translocation velocity~\cite{sakaue2007,sakaue2008,sakaue2010,saito2011,rowghanian2011}. However, this idea has not been quantitatively verified, since most of the studies have considered the asymptotic limit $N_0\rightarrow\infty$, which is out of reach of experiments and numerical simulations. %~\cite{sakaue2007,sakaue2008,sakaue2010,saito2011,rowghanian2011}. 
Therefore, it is imperative to study the TP mechanism for finite $N_0$, which is the regime that is experimentally relevant and where numerical simulation data are available.

To this end,  in this work we adopt the TP formalism in context of the Brownian dynamics (BD) equation of motion for $s$ mentioned above, in which we introduce a time-dependent friction coefficient $\Gamma=\Gamma(t)$ that is determined by the TP equations. We introduce a TP formalism for finite chain lengths by incorporating the pore-polymer interactions to the TP equations. We solve the resulting Brownian dynamics -- tension propagation (BDTP) model at finite chain length $N_0$, and validate it through extensive comparisons with molecular dynamics (MD) simulations. We verify that the tension propagation mechanism dominates the dynamics of driven translocation. In addition, we show that the model quantitatively reproduces the numerical values of $\alpha$ in various regimes without any free parameters, explaining the diversity in $\alpha$ as a finite chain length effect. Finally, we address the recent theoretical disagreement between the constant-velocity TP theory of Refs.~\cite{sakaue2007,sakaue2008,sakaue2010,saito2011} and the constant-flux TP theory of Ref.~\cite{rowghanian2011} and show that at the asymptotic limit, $N_0\rightarrow\infty$, $\alpha$ approaches $\alpha=1+\nu$.

\section{Model}
 \subsection{General formulation}

 We introduce dimensionless units for length, force, time, velocity and friction as $\tilde{s}=s/a$,  $\tilde{f}\equiv fa/k_BT$, $\tilde{t}\equiv tk_BT/\eta a^2$, $\tilde{v}\equiv v \eta a / k_BT$ and $\tilde{\Gamma}=\Gamma/\eta$, where $\eta$ is the solvent friction per monomer. In these units, the BD equation reads
\begin{equation}
\tilde{\Gamma}(\tilde{t})\frac{d\tilde{s}}{d\tilde{t}}=(1-\gamma')\left[ \frac{1}{N_0-\tilde{s}} -\frac{1}{\tilde{s}} \right] + \tilde{f} + \tilde{\zeta}(t) \equiv \tilde{f}_\mathrm{tot}, \label{eq:motion_dimless}
\end{equation}
where, for simplicity, we have assumed that the pore length $l_p=a$. Generalization of Eq.~(\ref{eq:motion_dimless}) to different pore lengths is straightforward (see, e.g., Ref~\cite{gauthier2008a}). Eq.~(\ref{eq:motion_dimless}) is, of course, approximative rather than rigorously exact. It contains two approximations, which we will show to be valid by quantitative comparison with MD simulations. First, we postulate that the friction $\tilde{\Gamma}(\tilde{t})$ is determined by TP on the {\it cis} side subchain. While there is no conclusive {\it a priori} reason to neglect the non-equilibrium effects of the {\it trans} side subchain, we will show that those effects are negligibly small in the experimentally and computationally relevant regimes. Second, we note that Eq.~(\ref{eq:motion_dimless}) includes the entropic force, whose form is strictly valid only for small driving forces $\tilde{f}$, when the translocation time $\tilde{\tau}$ is comparable to the Rouse relaxation time~\cite{rouse1953,chuang2001,dehaan2010}. However, for larger $\tilde{f}$, the average contribution of the entropic force to the total force $\tilde{f}_\mathrm{tot}$ is very small (see results below for discussion). Therefore, even for large forces, the model will be shown to give excellent agreement with MD simulations.

\begin{SCfigure}
%\centering
\includegraphics[width=0.48\columnwidth]{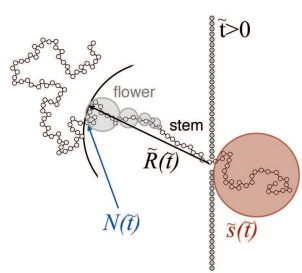}
\caption{(Color online) A snapshot of a translocating polymer in a stem-flower configuration. A tension front at $\tilde{x}=-\tilde{R}$ (black arc) divides the chain into moving and nonmoving parts, with the last moving monomer labeled as $N$. The number of translocated monomers is $\tilde{s}$. }
\label{fig:configuration}
\end{SCfigure}

The effective friction $\tilde{\Gamma}(\tilde{t})$ actually consists of two contributions. The first one is the drag force of the {\it cis} side subchain that is solved with the TP formalism. The other one is the frictional interaction between the pore and the polymer.  Formally, we can write $\tilde{\Gamma}$ as the sum of the {\it cis} side subchain and pore frictions, $\tilde{\Gamma}(\tilde{t})=\tilde{\eta}_\mathrm{cis}(\tilde{t})+\tilde{\eta}_p$. While for $N_0\rightarrow\infty$ the first term dominates, for finite $N_0$ the pore friction can significantly affect the translocation dynamics. We will come back to this issue later, but let us first look at how the time-dependent part of the friction can be determined from the TP formalism. In the special case of extremely large driving force, one can find $\tilde{\eta}_\mathrm{cis}$ directly from the TP equations. More generally, however, it is easier to derive the velocity of the monomers at the pore entrance, $\tilde{v}_0$. In such a case, the effective friction is in a natural way defined as
\begin{equation}
\tilde{\Gamma}(\tilde{t})= \frac{\tilde{f}_\mathrm{tot}}{\tilde{\sigma}_0(\tilde{t})\tilde{v}_0(\tilde{t})}, \label{eq:effective_friction}
\end{equation}
where $\tilde{\sigma}_0$ is the line density of monomers near the pore and $\tilde{\sigma}_0\tilde{v}_0\equiv d\tilde{s}/d\tilde{t}$ is the flux of monomers through the pore entrance. In either case, determining $\tilde{\Gamma}(\tilde{t})$ essentially reduces to calculating the number of moving monomers, whose combined drag force then constitutes the time-dependent part of the friction. As the driving force is applied, the chain begins to move in stages, with the segments closest to the pore being set into motion first. A close analogue is a coil of rope pulled from one end, which first uncoils before starting to move as a whole. To keep track of the moving part of the chain, one defines a {\it tension front}, which divides the chain into the moving part that is under tension, and the nonmoving part outside the front (see Fig.~\ref{fig:configuration}). The front is located at $\tilde{x}=-\tilde{R}(\tilde{t})$, and propagates in time as parts of the chain further away from the pore are set in motion. The last monomer within the tension front is labeled as $N(\tilde{t})$. Using the TP formalism, one can derive an equation of motion for the tension front, using either $\tilde{R}$ or $N$ as the dynamical variable. The details of this calculation can be found in the Appendix.

\subsection{Different regimes}

Depending on the magnitude of the driving force, the equation of motion for the tension front attains a slightly different form. In the simplest case, when the driving force is very large compared to temperature and chain length, $\tilde{f} \gtrsim N_0^\nu$, the moving part of the chain is almost completely straight. In this {\it strong stretching} (SS) regime, the equation of motion is~\cite{derivation_ss}
\begin{equation}
\frac{dN}{d\tilde{t}}=\frac{\tilde{f}_\mathrm{tot}}{\tilde{\Gamma}(\tilde{t}) \left(1-\nu A_\nu N^{\nu-1} \right)}.\label{eq:tp_ss}
\end{equation}
Here, $\tilde{\Gamma}(\tilde{t})=N(\tilde{t})-\tilde{s}(\tilde{t})+\tilde{\eta}_p$, with $N(\tilde{t})-\tilde{s}(\tilde{t})$ being the number of moving monomers on the {\it cis} side. The Flory exponent $\nu$ and the prefactor $A_\nu$ are related to the end-to-end distance of the polymer, $\tilde{R}_\mathrm{ee}=A_\nu N_0^\nu$. In the SS approximation, Eq.~(\ref{eq:motion_dimless}) is solved simultaneously with Eq.~(\ref{eq:tp_ss}), using $\tilde{\Gamma}$ from Eq.~(\ref{eq:tp_ss}) as a input in Eq.~(\ref{eq:motion_dimless}), and vice versa for $\tilde{s}$. 

For slightly smaller driving forces, $1 \lesssim \tilde{f} \lesssim N_0^\nu$, the force is not sufficient to completely straighten the chain. Due to thermal fluctuations, a flower-shaped tail develops (see Fig.~\ref{fig:configuration}). In this {\it stem-flower} (SF) regime, the line density and and velocity of the monomers are not constant in space. Therefore, one also has to solve the density $\tilde{\sigma}_R$ and the velocity $\tilde{v}_R$ near the tension front. As a result, one gets a system of equations,
\begin{align}
&\frac{d\tilde{R}}{d\tilde{t}}=\tilde{v}_{R} \left[ \frac{1}{\nu} A_\nu^{-1/\nu} \tilde{\sigma}_{R}^{-1} \tilde{R}^{1/\nu -1} \right]^{-1}, \label{eq:motionR}\\
&\tilde{\sigma}_R^{1/(\nu-1)}= \frac{\tilde{v}_0\tilde{R}}{\nu b \tanh(b)}\ln\left[\cosh\left(b\frac{\tilde{\sigma}_R^{\nu/(1-\nu)}}{\tilde{R}}\right)\right], \label{eq:sigmaR}\\
&\tilde{v}_R=\tilde{v}_0\frac{\tanh \left(b\tilde{\sigma}_R^{\nu/(1-\nu)} /\tilde{R} \right) }{\tanh(b)}, \label{eq:vR}\\
&\tilde{v}_0\tilde{R}\frac{\ln [\cosh (b)]}{b\tanh(b)}=\left[ \tilde{f}_\mathrm{tot}-\tilde{\eta}_p\tilde{v}_0  \right] + \nu -1, \label{eq:v0_sf}
\end{align}
that can be solved numerically for $\tilde{v}_0$. Here, $b$ is a (fixed) dimensionless parameter related to the spatial dependence of the velocity, and ensures global conservation of mass (see Appendix A). In the SF regime, $\tilde{\sigma}_0=1$, since the stem close to the pore is in a single-file configuration. The effective friction is given by Eq.~(\ref{eq:effective_friction}).

Finally, in the regime where the force insufficient to straighten even a small part of the chain, $\tilde{f} \lesssim N_0^{-\nu}$, the chain adopts a trumpet-like shape. In this {\it trumpet} (TR) regime, the dynamics can be described by Eqs.~(\ref{eq:motionR})--(\ref{eq:vR}), with the velocity $\tilde{v}_0$ and density $\tilde{\sigma}_0$ given by
\begin{align}
&\tilde{v}_0\tilde{R}\frac{\ln [\cosh (b)]}{b\tanh(b)}=\nu \left[ \tilde{f}_\mathrm{tot}-\tilde{\eta}_p\tilde{v}_0  \right]^{1/\nu}, \label{eq:v0_tr}\\
&\tilde{\sigma}_0= \left[ \tilde{f}_\mathrm{tot}-\tilde{\eta}_p\tilde{v}_0  \right]^{1-1/\nu}.
\end{align}

\subsection{Pore friction}

The time evolution of the tension front ($\tilde{R}$ or $N$) gives the contribution of the {\it cis} side subchain to the friction $\tilde{\Gamma}$. To complete the BDTP model, we still need to determine the pore friction $\eta_p$. In general, $\eta_p$ is a complicated function of the pore geometry, but here we restrict our study to the geometries used in our benchmark MD simulations. In order to fix $\eta_p$, we examine the waiting time per monomer $w(\tilde{s})$, defined as the time that the individual monomer spends inside the pore. With $\tilde{f}$ sufficiently large, $\tilde{w}\propto \tilde{\Gamma}/\tilde{f}$. For small $\tilde{s}$, the friction $\tilde{\Gamma}$ is mostly determined by $\tilde{\eta}_p$. Therefore, by comparing the $w(\tilde{s})$ of the BDTP model with MD simulations  of Refs.~\cite{huopaniemi2006,luo2009,kaifu_private} for the first few monomers,we have measured $\eta_p^\mathrm{3D}\approx 5$ and $\eta_p^\mathrm{2D}\approx 4$ for the respective pore geometries. It should be noted that $\eta_p$ is fitted only once, as opposed to being done separately for each combination of $\tilde{f},\eta$, etc. Thus, $\eta_p$ is {\it not} a freely adjustable parameter.

\section{Results and discussion}

\begin{figure}
\includegraphics[width=0.75\columnwidth]{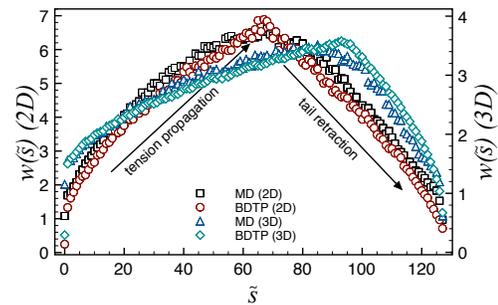}
\caption{(Color online) Comparison of waiting times $w$ in both 2D and 3D for MD and the BDTP model. The agreement of the BDTP model with MD simulations is excellent, and reveals the two stages of translocation: the tension propagation stage of increasing $w(\tilde{s})$ and the tail retraction stage characterized by decreasing $w(\tilde{s})$. The parameters used were the same for both MD and BDTP ($N_0=128$, $f=5$, $k_BT=1.2$, $\eta=0.7$). The 3D MD results are from~\cite{kaifu_private}.}
\label{fig:wt_N128}
\end{figure}

First, we validate the BDTP model through quantitative comparisons with MD simulations. In Fig.~\ref{fig:wt_N128}, we compare the waiting time $w(\tilde{s})$, which is the most important and sensitive measure of the translocation dynamics. As is shown, the match between BDTP and MD is almost exact. We stress that this agreement tells that the translocation dynamics is reproduced correctly at the most fundamental level and that such an agreement is a vital requirement for any correct theoretical model. The comparison also reveals an extremely lucid picture of the translocation process: first, as tension propagates along the chain, the effective friction increases and translocation slows down. In the second stage, the number of dragged monomers is reduced as the tail retracts, and translocation speeds up. 

\begin{figure}
\includegraphics[width=0.75\columnwidth]{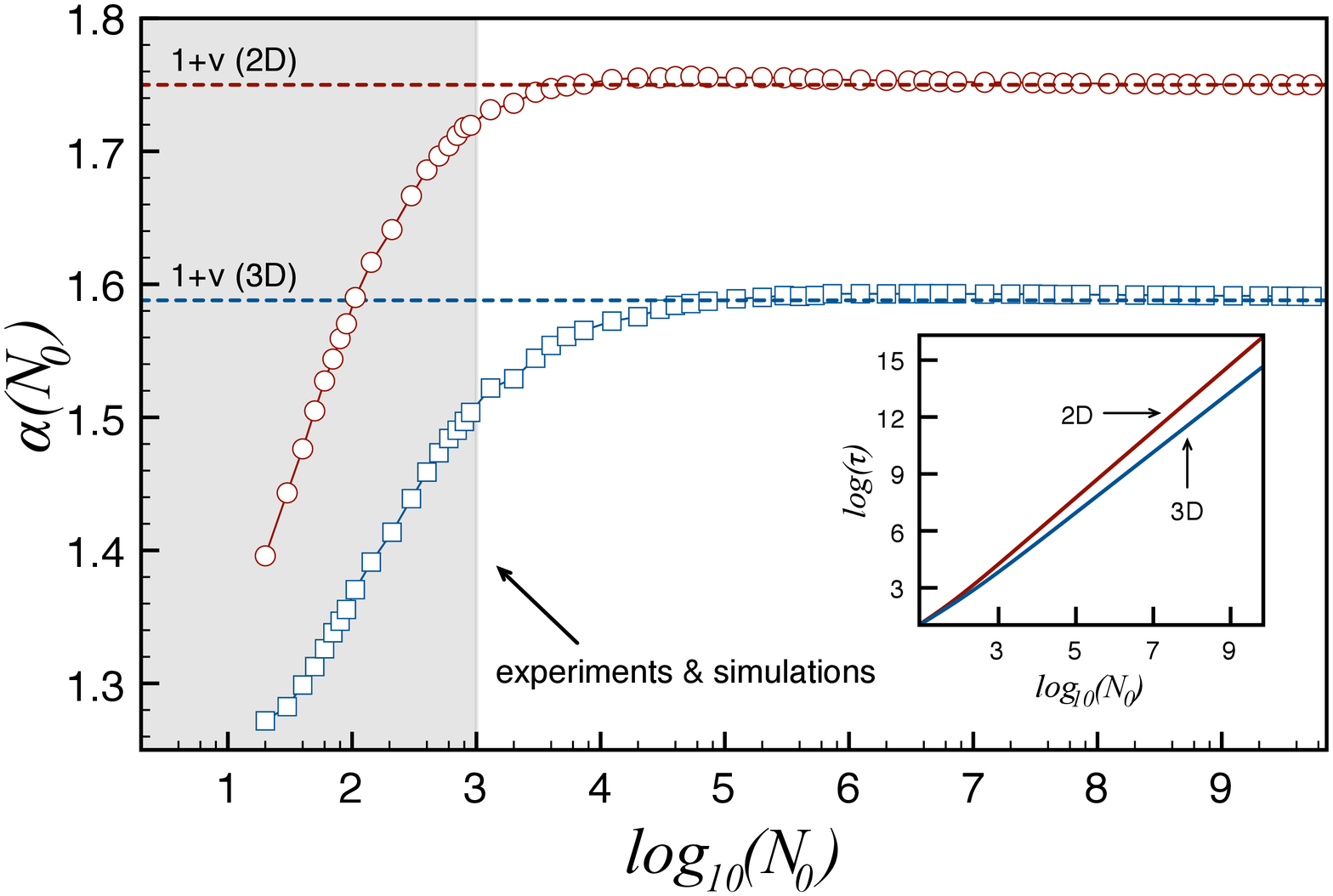}
\caption{(Color online) The effective exponent $\alpha(N_0)\equiv\frac{d\ln \tau}{d\ln N_0}$ in 2D (circles) and 3D (squares) as a function of $N_0$ from the BDTP model, showing the extremely slow approach to the asymptotic limit $\alpha=1+\nu$. Most of experimental and simulation studies in the literature involve chain lengths of $N_0 \lesssim 10^3$ (shaded region), being clearly in the finite chain length regime. The inset shows the raw data $\tau(N_0)$. Model parameters are the same as in Fig.~\ref{fig:wt_N128}.
}
\label{fig:alpha_asym}
\end{figure}

\begin{table}
\caption{Values of $\alpha$ ($\tau\sim N_0^\alpha$) from the BDTP model as compared to the corresponding values from MD simulations.}
\label{tb:alphas}
\begin{tabular} { l l l }
$\alpha$ (BDTP) &$\alpha$ (MD) & Dimension and parameter values \\
\hline
\hline
& & 2D, $T=1.2$, Ref.~\cite{huopaniemi2006} \\
  $1.51 \pm  0.02$ &  $1.50 \pm 0.01$ & $f=5.0$, $\gamma=0.7$, $20 \leq N_0 \leq 70$ \\
  $1.71 \pm 0.02$ &  $1.69 \pm 0.04$  & $f=5.0$, $\gamma=0.7$, $500 \leq N_0 \leq 800$ \\
  $1.52 \pm 0.02$ &  $1.50 \pm 0.02$  & $f=2.4$, $\gamma=0.7$, $20 \leq N_0 \leq 70$ \\
  $1.71 \pm 0.02$ &  $1.65 \pm 0.04$  & $f=2.4$, $\gamma=0.7$, $500 \leq N_0 \leq 800$ \\
  $1.66 \pm  0.02$ &  $1.64 \pm 0.01$ & $f=5.0$, $\gamma=3.0$, $20 \leq N_0 \leq 70$ \\
  $1.71 \pm 0.02$ &  $1.67 \pm 0.03$  & $f=5.0$, $\gamma=3.0$, $500 \leq N_0 \leq 800$ \\
\hline
& & 3D, $T=1.2$, Ref.~\cite{luo2009} \\
  $1.59 \pm  0.02$ &  $1.58 \pm 0.03$ & $f=0.5$, $\gamma=0.7$, $16 \leq N_0 \leq 128$ \\
  $1.35 \pm  0.02$ &  $1.37 \pm 0.05$ & $f=5.0$, $\gamma=0.7$, $16 \leq N_0 \leq 256$ \\
  $1.34 \pm  0.02$ &  $1.37 \pm 0.02$ & $f=10.0$, $\gamma=0.7$, $16 \leq N_0 \leq 256$ \\
\hline
& & 3D, $T=1.2$, Ref.~\cite{luo2008} \\
  $1.41 \pm  0.01$ &  $1.42 \pm 0.01$ & $f=5.0$, $\gamma=0.7$, $40 \leq N_0 \leq 800$ \\
  $1.39 \pm  0.02$ &  $1.41 \pm 0.01$ & $f=5.0$, $\gamma=0.7$, $64 \leq N_0 \leq 256$ \\
\hline
& & 3D, $T=1.0$, Ref.~\cite{lehtola2008} \\
  $1.46 \pm  0.02$ &  $1.47 \pm 0.05$ & $f=3.0$, $\gamma=11.7$, $70 \leq N_0 \leq 200$ \\
  $1.49 \pm  0.02$ &  $1.50 \pm 0.01$ & $f=30.0$, $\gamma=11.7$, $200 \leq N_0 \leq 800$ \\
\hline
\end{tabular}
\end{table}

Next, we compare the exponents $\alpha$ obtained from the BDTP model with the corresponding numerical values from MD simulations. The parameter range has been chosen to span the TR, SF and SS regimes, and to cover both short and long chain regimes in both 2D and 3D. The results are shown in Table I. The diversity of $\alpha$ in these regimes is evident, yet in all of them, the BDTP model is accurate to three significant numbers within the margin of error. This clearly shows, that while the values of $\alpha$ depend on several parameters, they all share a common physical basis: non-equilibrium tension propagation on the {\it cis} side subchain. Then why is the exponent $\alpha$ not universal? The answer lies in the chain length regimes studied both in experiments and simulations. Typically, $N_0\lesssim 10^3$. However, in this regime $\alpha$ is not independent of the chain length! As shown in Fig.~\ref{fig:alpha_asym}, $\alpha$ retains a fairly strong dependence on $N_0$ up to $N_0 \approx 10^4$ in 2D and $N_0 \approx 10^5$ in 3D. Therefore, {\it the observed scatter in $\alpha$ is a finite chain length effect}, a fact that has been mostly ignored in the literature.

Two additional remarks about the results of Table I are in order. First, the effect of the entropic term in Eq.~(\ref{eq:motion_dimless}) on $\alpha$ is extremely small. To show this, we solved Eq.~(\ref{eq:motion_dimless}) also without the entropic term. The results match exactly with those given in Table I, except for the low force case $f=0.5$, $T=1.2$ in 3D, where, without the entropic term, $\alpha=1.56$ instead of 1.59. Second, regarding $f$ and $N_0$, the BDTP model gives two general trends for $\alpha$: i) for a fixed $f$, $\alpha$ increases with $N_0$ (as shown in Fig.~\ref{fig:alpha_asym}) and ii) for a fixed $N_0$, $\alpha$ decreases with $f$, as shown in Table I for 2D and 3D (Refs.~\cite{huopaniemi2006,luo2009}). For $f/k_BT \ll 1$, this trend is consistent with the value of $\alpha$ in absence of $f$~\cite{luo2008}. For $f/k_BT \gg 1$, $\alpha$ is almost independent of $f$. Therefore, the increase of $\alpha$ in the last two lines of Table I (Ref.~\cite{lehtola2008}) is in fact due to increase in $N_0$, not in $f$.

Finally, we have estimated the asymptotic value of $\alpha$ in the SF regime by solving the BDTP model up to $N_0=10^{10}$ (Fig.~\ref{fig:alpha_asym}). In 2D, the numerical estimate is $\alpha^\mathrm{2D}_\infty \approx 1.750\pm 0.001$, for $N_0\gtrsim 10^9$, and, in 3D, $\alpha^\mathrm{3D}_\infty \approx 1.590\pm 0.002$, for $N_0\gtrsim 10^9$. In both cases, we recover the value $1+\nu$ as predicted in Ref.~\cite{rowghanian2011} with the constant-flux TP theory, and recently also using a different approach~\cite{dubbeldam2011}. However, the value is different from Sakaue's original prediction of $\frac{1+\nu+2\nu^2}{1+\nu}$~\cite{sakaue2007,sakaue2010,saito2011}. The reason for the different predictions is that in both Refs.~\cite{rowghanian2011, dubbeldam2011}, and in our model, the number of monomers is globally conserved, whereas in Refs.~\cite{sakaue2007,sakaue2010,saito2011} the conservation is guaranteed only locally in the neighborhood of $\tilde{x}=-\tilde{R}$. Therefore, asymptotically, $\alpha_\infty = 1+\nu$ in both 2D and 3D, also in agreement with the prediction of Ref.~\cite{kantor2004}. 

\section{Conclusions}

To summarize, we have introduced a new theoretical model of driven polymer translocation that has only two degrees of freedom and no free parameters. The model gives near-exact agreement with high-accuracy molecular dynamics simulations in a wide range of parameters.  Our study shows that the dynamics of driven translocation is dictated by non-equilibrium tension propagation on the {\it cis} side subchain.  The model also reveals that the majority of experiments and simulations in the literature are performed in the regime, where finite chain length effects have significant impact on the translocation dynamics. Although mostly overlooked in the literature, this is an important observation, since the finite chain length effects persists for chain lengths of at least several tens of thousands of monomers. Therefore, in most studies of polymer translocation, finite chain length effects cannot be neglected. This fact is also vital for the theoretical study of driven polymer translocation.

\acknowledgments

This work has been supported in part by the Academy of Finland through its COMP Center of Excellence and Transpoly Consortium grant. TI acknowledges the financial support of FICS and TES. The authors also wish to thank CSC, the Finnish IT center for science, for allocation of computer resources.

\appendix

\section{Derivation of the tension propagation equations}

For convenience, we use dimensionless units  denoted by the tilde symbol as $\tilde{X}\equiv X/X_u$, with the unit of length $a_u\equiv a$, force $f_u\equiv k_BT/a$, time $t_u\equiv \eta a^2 /k_BT$, velocity $v_u\equiv a/t_u$ and friction $\eta_u\equiv \eta$, where $\eta$ is the solvent friction per monomer. The tension propagation (TP) formalism is derived for $N_0\rightarrow\infty$ in Refs.~\cite{sakaue2007,sakaue2008,sakaue2010,saito2011}. Here, we derive the formalism for finite chain lengths by including the explicit  pore-polymer interactions through the pore friction $\tilde{\eta}_p$, and the spatially dependent velocity profile for the polymer chain. These will be discussed below.

To begin, we note that as the driving force at the pore is applied, the chain starts to move in stages, as tension propagates along the backbone. At time $\tilde{t}$, when $\tilde{s}(\tilde{t})$ monomers have translocated, $N(\tilde{t})-\tilde{s}(\tilde{t})$ monomers on the {\it cis} side are under tension and moving towards the pore at velocity $\tilde{v}(\tilde{x},\tilde{t})$ and, the remaining $N_0 - N(\tilde{t})$ monomers are at rest. The moving and unmoving domains are separated by a tension front at $\tilde{x}=-\tilde{R}$. The subchain between $-\tilde{R} \le\tilde{x}\le 0$ is deformed under the tension and can be regarded as a self-avoiding walk following the Pincus blob description~\cite{pincus1976}. The subchain adopts a configuration of increasing blob radii $\tilde{\xi}(\tilde{x})$, with the blob closest to the front having radius $\tilde{\xi}_R \equiv \tilde{\xi}(-\tilde{R}+\tilde{\xi}_R)$. At length scales shorter than the blob size $\tilde{\xi}(\tilde{x})=1/\tilde{f}(\tilde{x})$, the chain behaves as if undisturbed by the driving force, scaling as $\tilde{\xi}=g^\nu$, where $g$ is the number of monomers inside the blob and $\nu$ is the Flory exponent. This gives the relation $\tilde{\sigma}=g/\tilde{\xi}=\tilde{\xi}^{1/\nu-1}$ for the monomer line density  $\tilde{\sigma}$. By definition, $\tilde{\sigma}$ satisfies
$\int_{-\tilde{R}(\tilde{t})}^{0}\tilde{\sigma}(\tilde{x}',\tilde{t})d\tilde{x}'=N(\tilde{t})-\tilde{s}(\tilde{t}).$ To solve $\tilde{\xi}(\tilde{x})$, we require the local force balance between the driving force and the drag force at $\tilde{x}$:
\begin{equation}
\int_{-\tilde{R}(\tilde{t})}^{\tilde{x}}\tilde{v}(\tilde{x}',\tilde{t})[\tilde{\xi}(\tilde{x}',\tilde{t})]^{1/\nu-1}d\tilde{x}'=\tilde{f}(\tilde{x},\tilde{t})=\tilde{\xi}(\tilde{x},\tilde{t})^{-1}.\label{eq:Alocal_fb}
\end{equation}
In addition, there is a balance between the driving force at the pore entrance, $\tilde{f}_0\equiv \tilde{f}(0)$ and the total drag force of the {\it cis} side subchain. This global force balance can be enforced by substituting $\tilde{x}=0$ to Eq.~(\ref{eq:Alocal_fb}):
\begin{equation}
 \int_{-\tilde{R}(\tilde{t})}^{0}\tilde{v}(\tilde{x}',\tilde{t})\tilde{\sigma}(\tilde{x}',\tilde{t})d\tilde{x}'=\tilde{f}_0 \label{eq:Aglobal_fb}
\end{equation}

The time evolution of the tension front $\tilde{R}$ obeys the equation of conservation of monomers,
\begin{equation}
\tilde{\sigma}_R(\tilde{t})\left[ \frac{d\tilde{R}(\tilde{t})}{d\tilde{t}}+\tilde{v}_R(\tilde{t}) \right] = \frac{dN(\tilde{t})}{d\tilde{t}}.\label{eq:Amono_conserv}
\end{equation}
Here we employ the short-hand notation $\tilde{\sigma}_R(\tilde{t})\equiv \tilde{\sigma}(-\tilde{R}+\tilde{\xi}_R,\tilde{t})$ and $\tilde{v}_R(\tilde{t})=\tilde{v}(-\tilde{R}+\tilde{\xi}_R,\tilde{t})$. Furthermore, since the monomers outside the tension front are on the average immobile, the location of the tension front $\tilde{R}$ is given by the equilibrium end-to-end distance of the subchain consisting of the first $N$ monomers~\cite{sakaue2007}:
\begin{equation}
\tilde{R}(\tilde{t})=A_\nu N(\tilde{t})^{\nu}.\label{eq:Aclosure}
\end{equation}
Here, $A_\nu$ is a model-dependent prefactor for a chain with one end tethered to a wall. For the ideal chain, $A_\nu=1$ and, for the self-avoiding chain we have measured $A_\nu\approx 1.16\pm 0.05$ in 2D and $1.15\pm 0.03$ in 3D from MD simulations using the Kremer-Grest model~\cite{kremer-grest} with typical parameter values for the chain. Details of the simulation are explained in Appendix B.

To couple the TP model with the effective friction of the Brownian dynamics equation of $\tilde{s}$, we write down the conservation of monomers at the pore entrance,
$\frac{d\tilde{s}(\tilde{t})}{d\tilde{t}}=\tilde{\sigma}_0(\tilde{t})\tilde{v}_0(\tilde{t}),$
which also defines
$\tilde{\Gamma}(\tilde{t})= \frac{\tilde{f}_\mathrm{tot}}{\tilde{\sigma}_0(\tilde{t})\tilde{v}_0(\tilde{t})}, $
where $\tilde{\sigma}_0(\tilde{t})=\tilde{\sigma}(0,\tilde{t})$ and $\tilde{v}_0(\tilde{t})=\tilde{v}(0,\tilde{t})$. Finally, we define the explicit relationship between the total driving force $\tilde{f}_\mathrm{tot}$ and the force at the pore entrance, $\tilde{f}_0$. For finite $N_0$, we need to take into account the pore-polymer interactions by introducing the pore friction coefficient $\tilde{\eta}_p=\eta_p/\eta$. This defines the relationship between $\tilde{f}_\mathrm{tot}$ and $\tilde{f}_0$ as
$\tilde{f}_\mathrm{tot} -\tilde{\eta}_p\tilde{v}_0(\tilde{t}) = \tilde{f}_0.$
Obviously, for an ideally frictionless pore, $\tilde{\eta}_p=0$ and $\tilde{f}_\mathrm{tot} = \tilde{f}_0$. For finite $N_0$ and $\tilde{\eta}_p$, the overall effect of $\tilde{\eta}_p$ is to make the translocation time $\tau$ less sensitive to $N_0$, i.e., decrease $\alpha$.

{\it Strong stretching (SS) regime.} Let us first consider the case of a strong driving force so that the moving part of the chain is almost completely straight. This strong stretching regime is realized when $\tilde{f}_0 \gtrsim N_0^\nu$~\cite{sakaue2007}. In the SS regime, the line density of monomers in the moving domain is constant $\tilde{\sigma}(\tilde{x})=\tilde{\sigma}^*\approx 1$ and correspondingly $\tilde{\xi}(\tilde{x})=\tilde{\xi}^*\approx1$. As a first approximation, one may assume that the segment length $a$ remains unchanged. Although a more detailed treatment is possible~\cite{saito2011}, this is a reasonable approximation in the relevant range of forces. From these assumptions, one immediately obtains a constant velocity profile for the monomers in the moving domain: $\tilde{v}(\tilde{x},\tilde{t})=\tilde{v}_0(\tilde{t})\Theta(\tilde{x}+\tilde{R})$, where $\Theta(\tilde{x})$ is the Heaviside step function. In particular, $\tilde{v}_0(\tilde{t})=\tilde{v}_R(\tilde{t})~\forall \tilde{t}$. In the SS regime, one obtains the TP law
\begin{equation}
\frac{dN}{d\tilde{t}}=\frac{\tilde{f}_\mathrm{tot}}{\tilde{\Gamma}(\tilde{t}) \left(1-\nu A_\nu N^{\nu-1} \right)},\label{eq:Atp_ss}
\end{equation}
with $\tilde{\Gamma}(\tilde{t})=N(\tilde{t})-\tilde{s}(\tilde{t})+\tilde{\eta}_p$. Note that Eq.~(\ref{eq:Atp_ss}) is a generalization of the TP law of Refs.~\cite{sakaue2008, saito2011} to pore-driven translocation with an explicit pore friction $\eta_p$.

{\it Trumpet (TR) regime.} In the opposite limit of extremely weak driving force ($\tilde{f}_0 \lesssim N_0^{-\nu}$), the whole chain can be considered to be in equilibrium, and one recovers the Rouse-type friction of the subchain. However, such small forces are rarely realized in simulations or experiments. Indeed, if the force is only slightly larger, $N_0^{-\nu} \lesssim \tilde{f}_0 \lesssim 1$~, the chain adopts a configuration resembling a trumpet, where the blob radius $\tilde{\xi}$ increases as one moves further away from the pore \cite{sakaue2007,saito2011}. This leads to $\tilde{\sigma}_0(\tilde{t})<\tilde{\sigma}_R(\tilde{t}) ~\forall \tilde{t}$. In Refs.~\cite{sakaue2007,sakaue2010,saito2011}, it is assumed that the velocity profile is constant in the TR regime, similarly to the SS regime. This assumption, however, leads to a contradiction. By integrating Eq.~(\ref{eq:Amono_conserv}) over the whole translocation process from $\tilde{t}=0$ to $\tilde{t}=\tilde{\tau}$, it follows that $\tilde{s}(\tilde\tau)<N_0$. This is an obvious contradiction, a point that was also raised in Ref.~\cite{rowghanian2011}. The only way to remove the contradiction is to relax the constant-velocity assumption so that $\tilde{v}_0(\tilde{t})\geq \tilde{v}_R(\tilde{t})$. It should be noted that the constant-flux approximation of Ref.~\cite{rowghanian2011} is also not valid in the short chain regime, where $d\tilde{R}/d\tilde{t}$ is of the order of $d\tilde{s}/d\tilde{t}$. To solve this problem, we have studied the velocity profile using MD simulations, and find that at least for $N_0 < 10^3$, the velocity profile is to good approximation given by  $\tilde{v}(\tilde{x},\tilde{t})=\tilde{v}_0(\tilde{t}) \frac{\tanh \left[ b\left( \tilde{x}/\tilde{R}+1\right)\right]}{\tanh(b)}$~\cite{velocitynote}. Here $b$ is a dimensionless parameter that controls the sharpness of the profile and is fixed by enforcing global conservation of monomers, i.e., requiring that $\tilde{s}(\tilde{\tau})=N_0$ and $\tilde{R}(\tilde{\tau})=0$. Because the approximate profile for $\tilde{v}(\tilde{x},\tilde{t})$ is not exact, $b$ has a weak dependence on chain length $N_0$ and is solved for each $N_0$ by numerical iteration until $\tilde{s}(\tilde{\tau})=N_0$ and $\tilde{R}(\tilde{\tau})=0$ is satisfied.

In the TR regime, the line density $\tilde{\sigma}$ of the monomers is not fixed, but is determined by Eq.~(\ref{eq:Alocal_fb}). The line density near the pore is determined  by the blob radius $\tilde{\xi}(0)=\tilde{f}_0^{-1}$. To calculate the line density at the boundary, one has to solve Eq.~(\ref{eq:Alocal_fb}) for $\tilde{x}\in \left[ -\tilde{R}, -\tilde{R}+\tilde{\xi}_R \right]$. The resulting implicit equation for $\tilde{\xi}_R$ is solved numerically:
\begin{equation}
\tilde{\xi}_R(\tilde{t})^{-1/\nu}= \frac{\tilde{v}_0(\tilde{t})\tilde{R}(\tilde{t})}{\nu b \tanh(b)}\ln\left[\cosh\left(b\frac{\tilde{\xi}_R(\tilde{t})}{\tilde{R}(\tilde{t})}\right)\right]. \label{eq:AxiR}
\end{equation}
The velocity $\tilde{v}_0(\tilde{t})$ can be solved from Eq.~(\ref{eq:Alocal_fb}) with $\tilde{x}=0$, which reduces into another implicit equation,
\begin{equation}
\tilde{v}_0(\tilde{t})\tilde{R}(\tilde{t})\frac{\ln [\cosh (b)]}{b\tanh(b)}=\nu \left[ \tilde{f}_\mathrm{tot}-\tilde{\eta}_p\tilde{v}_0(\tilde{t})  \right]^{1/\nu}. \label{eq:Av0_tr}
\end{equation}
The velocity near the boundary is given by 
\begin{equation}
\tilde{v}_R(\tilde{t})\equiv \tilde{v}(-\tilde{R}+\tilde{\xi}_R,\tilde{t})=\tilde{v}_0(\tilde{t})\frac{\tanh \left(b\tilde{\xi}_R /\tilde{R} \right) }{\tanh(b)}. \label{eq:AvR}
\end{equation}
The equation of motion for the tension front can be solved from Eqs.~(\ref{eq:Amono_conserv}) and (\ref{eq:Aclosure}) that give
\begin{equation}
\frac{d\tilde{R}(\tilde{t})}{d\tilde{t}}=\tilde{v}_{R}(\tilde{t})  \left[ \frac{1}{\nu} A_\nu^{-1/\nu} \tilde{\sigma}_{R}(\tilde{t})^{-1} \tilde{R}(\tilde{t})^{1/\nu -1} \right]^{-1}.  \label{eq:AmotionR}
\end{equation}
Finally, the time evolution of the tension front is given by numerically solving Eqs.~(\ref{eq:AxiR})--(\ref{eq:AmotionR}) with the initial condition $\tilde{R}(\tau_0)\approx \tilde{f}_0^{-1}$~\cite{sakaue2007}.

{\it Stem-flower (SF) regime.} In the intermediate regime, $1 \lesssim \tilde{f}_0 \lesssim N_0^\nu$, the chain assumes a shape consisting of a fully elongated stem between the pore and $\tilde{x}=-\tilde{r}$, followed by a trumpet-shaped flower for $-\tilde{R}\leq \tilde{x} < - \tilde{r}$. The analysis of the stem is similar to the SS regime, and the flower in turn follows the TR regime calculation. The two parts are connected via the boundary condition $\tilde{f}(-\tilde{r})=1$, which, together with Eq.~(\ref{eq:Alocal_fb}) evaluated at $-\tilde{r}$, can be used to eliminate $\tilde{r}$, giving the equation for $\tilde{v}_0(\tilde{t})$:
\begin{equation}
\tilde{v}_0(\tilde{t})\tilde{R}(\tilde{t})\frac{\ln [\cosh (b)]}{b\tanh(b)}=\left[ \tilde{f}_\mathrm{tot}-\tilde{\eta}_p\tilde{v}_0(\tilde{t})  \right] + \nu -1. \label{eq:Av0_sf}
\end{equation}
The blob radius and the velocity at the boundary are given by Eqs. (\ref{eq:AxiR}) and (\ref{eq:AvR}), respectively, and the time evolution of the front by Eq. (\ref{eq:Amono_conserv}). Note that Eqs. (\ref{eq:Av0_tr}) and (\ref{eq:Av0_sf}) ensure a smooth cross-over between the TR and SF regimes at $\tilde{f}_0=1$ and that the SF regime equations approach the SS regime Eq.~(\ref{eq:Atp_ss}) when $\tilde{f} \gg 1$ (as $\tilde{r}\rightarrow \tilde{R}$, $\tilde{\xi}_R\rightarrow 1$ and $b\rightarrow\infty$).  In practice, we solve Eqs.~(\ref{eq:Amono_conserv}) and (\ref{eq:AxiR})--(\ref{eq:Av0_sf}), choosing Eq.~(\ref{eq:Av0_tr}) over Eq.~(\ref{eq:Av0_sf}) if $\tilde{f}_0 < 1$, and vice versa.

\section{Molecular dynamics simulations}

The details of the molecular dynamics simulations that we have used for benchmarking the BDTP model are explained in this Section. In the MD simulations, the polymer chain is modeled as Lennard-Jones particles interconnected by nonlinear FENE springs. Excluded volume interaction between monomers is given by the short-range repulsive Lennard-Jones potential:
\begin{equation}
U_\mathrm{LJ}(r)=
\begin{cases} 4\epsilon \left[ \left(\frac{\sigma}{r}\right)^{12} -  \left(\frac{\sigma}{r}\right)^{6}\right] +\epsilon & \text{for } r\leq 2^{1/6}\sigma\\
0 & \text{for } r>2^{1/6}\sigma
\end{cases}
\end{equation}
Here, $r$ is the distance between monomers, $\sigma$ is the diameter of the monomer and $\epsilon$ is the depth of the potential well. Consecutive monomers are also connected by FENE springs with 
\begin{equation}
U_\mathrm{FENE}(r)=-\frac{1}{2}kR_0^2\ln (1-r^2/R_0^2),
\end{equation}
where $k$ is the FENE spring constant and $R_0$ is the maximum allowed separation between consecutive monomers. For the chain, we use the parameters $\epsilon=1$, $k=15$ and $R_0=2$. The main part of the wall is constructed using a repulsive external potential of the Lennard-Jones form $U_\mathrm{ext}=4\epsilon \left[ \left(\frac{\sigma}{x}\right)^{12} -  \left(\frac{\sigma}{x}\right)^{6}\right] +\epsilon$ for $x\leq 2^{1/6}\sigma$ and 0 otherwise. Here $x$ is the coordinate in the direction perpendicular to the wall, with $x<0$ signifying the {\it cis} side and $x>0$ the {\it trans} side. The neighborhood of the pore is constructed of immobile Lennard-Jones beads of size $\sigma$. All monomer-pore particle pairs have the same short-range repulsive LJ-interaction as described above. We have verified that using the simple external potential $U_\mathrm{ext}$ for the wall gives the same results (within statistical error) as using a wall made of monomers in fixed lattice sites, at least as long as the interaction between the wall and the polymer is purely repulsive.

Similarly to most of the molecular dynamics simulations in the literature, %~\cite{huopaniemi2006,lehtola2008,lehtola2009,bhatta2009,bhatta2010,luo2008,luo2009,metzler2010,dubbeldam2011,huopaniemi2007}, 
we take the surrounding solvent into account through frictional and random forces. Thus, each monomer is described by the Langevin equation of motion 
\begin{equation}
m \mathbf{\ddot{r}}_i=-\nabla (U_\mathrm{LJ}+U_\mathrm{FENE} + U_\mathrm{ext})-\eta \mathbf{v}_i +\zeta_i,\label{eq:AmotionMD}
\end{equation}
where $m$ is the monomer mass, $\eta$ is the friction coefficient, $\mathbf{v}_i$ is the monomer velocity, $-\nabla U_\mathrm{ext} \equiv f$ is the external force in the pore and $\zeta_i$ is the random force with the correlations $\langle  \zeta_i(t) \zeta_j(t') \rangle = 2\eta k_BT \delta_{i,j}\delta(t-t')$, where $k_BT$ is the thermal energy. Typically, we have used the parameter values $m=1$, $\eta=0.7$, $k_BT=1.2$. The equations of motion are solved with the BBK algorithm~\cite{BBK} with time step $\delta t = 0.005$. Initially, the polymer chain is placed on the {\it cis} side with the first monomer located at the pore entrance. Eq.~(\ref{eq:AmotionMD}) is then solved numerically while keeping the first monomer fixed until an uncorrelated initial configuration is generated. After that, the whole chain is allowed to evolve according to Eq.~(\ref{eq:AmotionMD}) until the chain escapes either to the {\it cis} or {\it trans} side. The latter is recorded as a successful translocation event. We average our MD results over at least 2000 such events.


\begin{thebibliography}{99}

\bibitem{albertsbook} B. Alberts et al, {\it Molecular Biology of the Cell, 5th Ed.} (Garland, New York) 2008.

\bibitem{meller2003} A. Meller, J. Phys. Condens. Matter {\bf 15}, R581 (2003).

\bibitem{kasi1996} J.J. Kasianowicz, E. Brandin, D. Branton and D.W. Deamer, Proc. Natl. Acad. Sci. {\bf 93}, 13770 (1996).

\bibitem{storm2005} A.J. Storm {\it et al}, Nano Lett. {\bf 5}, 1193 (2005).

\bibitem{milchev2011} A. Milchev, J. Phys. Condens. Matter {\bf 23}, 103101 (2011).

\bibitem{sung1996} W. Sung and P.J. Park, Phys. Rev. Lett. {\bf 77}, 783 (1996).

\bibitem{muthu1999} M. Muthukumar, J. Chem. Phys. {\bf 111}, 10371 (1999).

\bibitem{kantor2004} Y. Kantor and M. Kardar, Phys. Rev. E {\bf 69}, 021806 (2004).

\bibitem{dubbeldam2007} J.L.A. Dubbeldam, A. Milchev, V.G. Rostiashvili and T.A. Vilgis,  Europhysics Lett. {\bf 79}, 18002 (2007).

\bibitem{vocks2008} H. Vocks, D. Panja, G.T. Barkema and R.C. Ball, J. Phys. Condens. Matter {\bf 20}, 095224 (2008).

\bibitem{sakaue2007} T. Sakaue, Phys. Rev. E {\bf 76}, 021803 (2007).

\bibitem{sakaue2010} T. Sakaue, Phys. Rev. E {\bf 81}, 041808 (2010).

\bibitem{saito2011} T. Saito and T. Sakaue, arXiv:1103.0620v1 (2011).

\bibitem{sakaue2008} T. Sakaue, in proceedings {\it The 5th Workshop on Complex Systems}, AIP CP, 982, 508 (2008).

\bibitem{rowghanian2011} P. Rowghanian and A. Y. Grosberg, J. Phys. Chem. B (2011).

\bibitem{dubbeldam2011} J. L. A. Dubbeldam, V. G. Rostiashvili, A. Milchev and T. A. Vilgis, arXiv:1110.5763v1 (2011).

\bibitem{chuang2001} J. Chuang, Y. Kantor and M. Kardar, Phys. Rev. E {\bf 65}, 011802 (2001).

\bibitem{huopaniemi2006} I. Huopaniemi, K. Luo, T. Ala-Nissila and S.C. Ying, J. Chem. Phys. {\bf 125}, 124901 (2006).

\bibitem{gauthier2008a} M.G. Gauthier and G.W. Slater, J. Chem. Phys. {\bf 128}, 065103 (2008).

\bibitem{gauthier2008b} M.G. Gauthier and G.W. Slater, J. Chem. Phys. {\bf 128}, 205103 (2008).

\bibitem{lehtola2008} V.V. Lehtola, R.P. Linna and K. Kaski, Phys. Rev. E {\bf 78}, 061803 (2008).

\bibitem{lehtola2009} V. Lehtola, R.P. Linna and K. Kaski, Europhys. Lett. {\bf 85}, 58006 (2009).

\bibitem{bhatta2009} A. Bhattacharya, W.H. Morrison, K. Luo, T. Ala-Nissila, S.-C. Ying, A. Milchev and K. Binder, Eur. Phys. J. E {\bf 29}, 423-429 (2009).

\bibitem{bhatta2010} A. Bhattacharya and K. Binder, Phys. Rev. E {\bf 81}, 041804 (2010).

\bibitem{luo2008} K. Luo, S.T.T. Ollila, I. Huopaniemi, T. Ala-Nissila, P. Pomorski, M. Karttunen, S.-C. Ying and A. Bhattacharya, Phys. Rev. E {\bf 78} 050901(R) (2008).

\bibitem{luo2009} K. Luo, T. Ala-Nissila, S.-C. Ying and R. Metzler, Europhys. Lett. {\bf 88}, 68006 (2009).

\bibitem{metzler2010} R. Metzler and K. Luo, Eur. Phys. J. Special Topics {\bf 189}, 119 (2010).

\bibitem{huopaniemi2007} I. Huopaniemi, K. Luo, T. Ala-Nissila and S.C. Ying, Phys. Rev. E. {\bf 75}, 061912 (2007).

\bibitem{rouse1953} P.E. Rouse, J. Chem. Phys. {\bf 21}, 1272 (1953).

\bibitem{dehaan2010} H.W. de Haan and G.W. Slater, Phys. Rev. E {\bf 81}, 051802 (2010).

\bibitem{pincus1976} P. Pincus, Macromolecules {\bf 9}, 386 (1976).

\bibitem{derivation_ss} For the derivation of Eq.~(\ref{eq:tp_ss}), the interested reader may refer to the Appendix and Refs.~\cite{sakaue2007,sakaue2008,dubbeldam2011}.

\bibitem{kaifu_private} K. Luo, private communication.

\bibitem{kremer-grest} G.S. Grest and K. Kremer, Phys. Rev. A {\bf 33}, 3628 (1986).

\bibitem{velocitynote} It turns out that the exact functional form of the velocity profile is not crucial. We also implemented a piecewise linear approximation of the velocity profile, where the velocity changes linearly from 0 to $\tilde{v}_0$ within distance $\Delta\tilde{R}$ of $\tilde{x}=-\tilde{R}$. Compared to the hyperbolic tangent profile, the difference in the finite chain length results is of the order of the statistical error, and the asymptotic limit is unchanged. This indicates that the model's results are not very sensitive to the details of the velocity profile. It is important, however, that the profile is such that the number of monomers is globally conserved.

\bibitem{BBK} A. Br{\"u}nger, L. Brooks III and M. Karplus, Chem. Phys. Lett. \textbf{105}, 495 (1984).


\end{thebibliography}
\end{document}